# Multilayer spintronic neural networks with radio-frequency connections

Andrew Ross[1†], Nathan Leroux[1†], Arnaud de Riz[1], Danijela Marković[1], Dédalo Sanz-Hernández[1], Juan Trastoy[1], Paolo Bortolotti[1], Damien Querlioz[2], Leandro Martins[3], Luana Benetti[3], Marcel S. Claro[3], Pedro Anacleto[3], Alejandro Schulman[3], Thierry Taris[4], Jean-Baptiste Begueret[4], Sylvain Saïghi[4], Alex S. Jenkins[3], Ricardo Ferreira[3], Adrien F. Vincent[4], Alice Mizrahi[1*] and Julie Grollier[1*]

[1]Unité Mixte de Physique CNRS/Thales, CNRS, Université Paris Saclay, 91767, Palaiseau, France

[2]Université Paris-Saclay, CNRS, Centre de Nanosciences et de Nanotechnologies, 91120 Palaiseau, France

[3]International Iberian Nanotechnology Laboratory (INL), 4715-31 Braga, Portugal

[4]Laboratoire de l'Intégration du Matériau au Système, Univ. Bordeaux, Bordeaux INP, CNRS, France

[†] these two authors have contributed equally

[*] alice.mizrahi@cnrs-thales.fr, julie.grollier@cnrs-thales.fr

**Spintronic nano-synapses and nano-neurons perform complex cognitive computations with high accuracy thanks to their rich, reproducible and controllable magnetization dynamics[1]. These dynamical nanodevices could transform artificial intelligence hardware, provided that they implement state-of-the art deep neural networks[2]. However, there is today no scalable way to connect them in multilayers. Here we show that the flagship nano-components of spintronics, magnetic tunnel junctions, can be connected into multilayer neural networks where they implement both synapses and neurons thanks to their magnetization dynamics, and communicate by processing, transmitting and receiving radio frequency (RF) signals. We build a hardware spintronic neural network composed of nine magnetic tunnel junctions connected in two layers, and show that it natively classifies nonlinearly-separable RF inputs with an accuracy of 97.7%. Using physical simulations, we demonstrate that a large network of nanoscale junctions can achieve state-of the-art identification of drones from their RF transmissions, without digitization, and consuming only a few milliwatts, which is a gain of more than four orders of magnitude in power consumption compared to currently used techniques. This study lays the foundation for deep, dynamical, spintronic neural networks.**

The recent progress in Artificial Intelligence (AI) relies on the ability to train deep neural networks which create, layer after layer, more and more meaningful representations of the inputs[2]. In these models, neurons perform a weighted sum of the output of neurons in the previous layer, where each weight represents a synapse, and then applies a non-linear function to this sum. Neuromorphic computing seeks to reproduce this brain-inspired multi-layer architecture and operations on chip with nanoscale artificial synapses and neurons, in order to place vast amounts of memory at the closest point to computing elements, speed up computing and reduce the overall energy consumption by several orders of magnitude compared to current hardware processors[3].

Spintronics possesses essential qualities for this purpose[1] as its flagship component, the magnetic tunnel junction, has a high endurance and provides vast amounts of memory integrated in the CMOS process of major foundries[4,5]. In recent years, the key effects of spintronics, such as magnetization dynamics and spin-charge conversion, have been combined to produce a large number of different spintronic nano-synapses[6–15] and nano-neurons[16–25] with multiple functionalities. Notably, the rich dynamics of small ensembles of oscillating magnetic tunnel junctions has been leveraged to recognize spoken words with software-equivalent accuracy through brain-inspired processes such as transient dynamics or synchronization[16,26,27]. To fulfill their promise, however, these dynamic nanodevices must be, like in the brain and in state-of-the-art algorithms, densely connected into neural networks with a multilayer architecture of nonlinear neurons connected by synapses that perform a weighted sum over the neurons' outputs[2].

We demonstrate a novel architecture for dense multilayer neural networks, exemplified in Fig. 1a. It leverages the intrinsic dynamics and magnetoresistance of magnetic tunnel junctions so that these devices emulate both synapses and neurons through successive RF-to-DC and DC-to-RF conversions. A major advantage of this approach is that the resulting neural networks can natively process RF inputs in the wide band of frequencies covered by magnetic tunnel junctions (typically 5 MHz to 20 GHz[28]) and perform, e.g., a direct recognition of airborne RF signals, thus avoiding the energy-expensive digitalization step normally used to apply AI techniques to RF signals[29,30]. We describe the concept of such networks, demonstrate experimentally the operations of its building blocks and show that a hardware network of nine magnetic tunnel junctions classifies non-linearly separable RF inputs with excellent accuracy. We then demonstrate, through physical simulations, the state-of-the-art accuracy and low power consumption of a large-scale network performing RF signal classification from drone emissions, and calculate a gain of more than four orders of magnitude in power consumption for this task compared to standard techniques. The magnetic tunnel junctions that we use in this article are all identical with a diameter of 350 nm, with different applied magnetic fields setting them in the desired range of operation, and a stack and properties detailed in Methods.

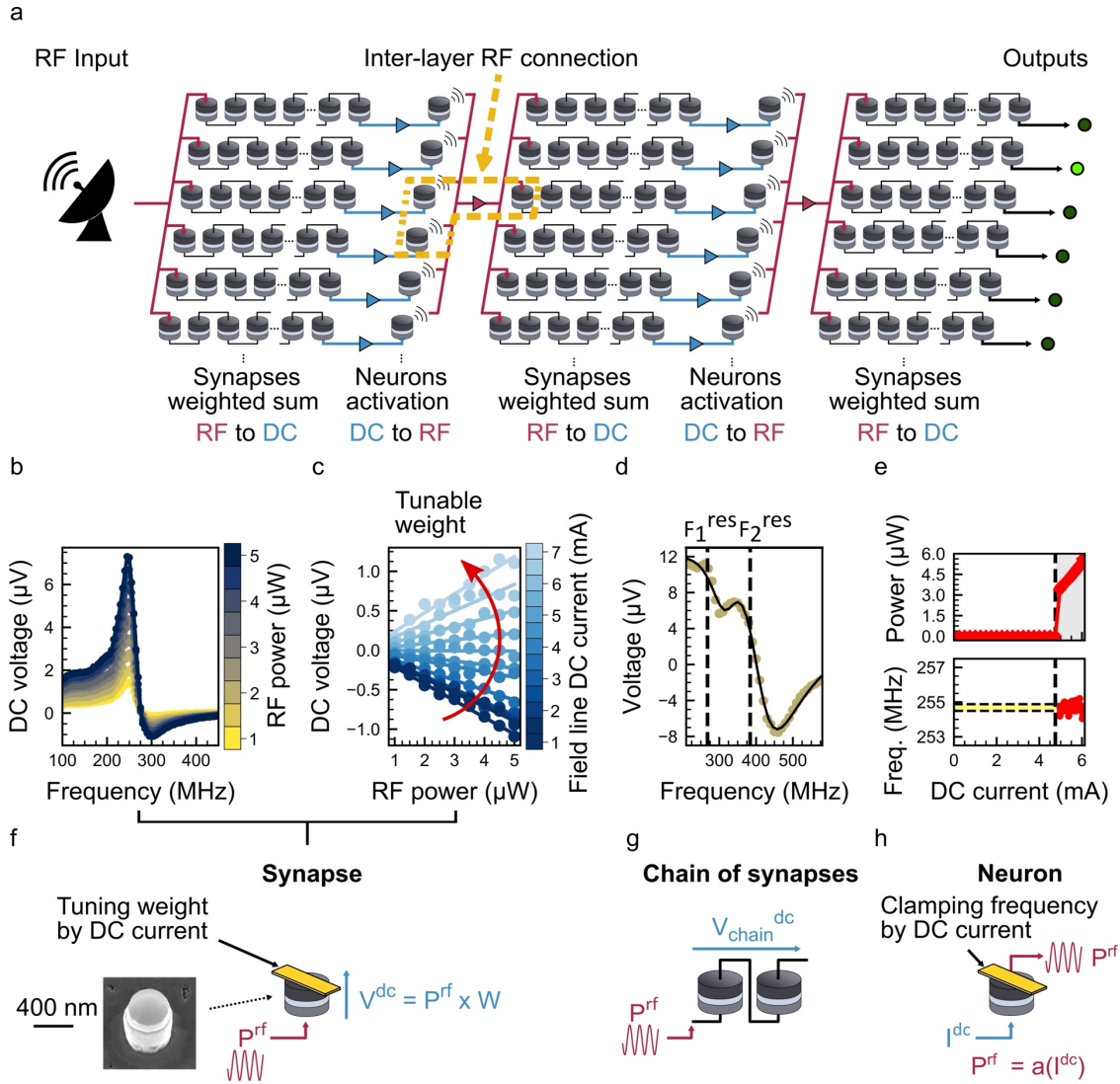

**Figure 1: Building blocks of multilayer RF/DC spintronic neural networks.** (a) Schematic of the proposed network. The cylinders represent magnetic tunnel junctions. RF and DC connections are shown in red and blue respectively. The triangles are amplifiers. The dashed yellow line highlights the neuron-to-synapse interconnection studied in Fig. 2. (b) Spin-diode response of a single synaptic junction: rectified DC voltage versus frequency of the input RF signal, for different input RF powers (color scale), under a 410 mT perpendicular magnetic field. (c) Synaptic multiplication: output DC voltage versus input RF power, at an input frequency of 285 MHz, for different synaptic weights (color scale: current in field line with 0.5 mA steps). (d) Spin-diode response of a chain of two synaptic junctions connected in series: rectified DC voltage versus frequency of the input signal, for an RF power of 7 µW, under a perpendicular field of 193 mT. The vertical dashed lines indicate the resonance frequencies of each synapse. The dots are experimental data and the solid line is a fit (combination of Lorentzian and anti-Lorentzian resonances). (e) Emitted RF power and frequency of a neuron, versus the DC current in the device, under a 415 mT perpendicular field. The vertical dashed line is the threshold current below which there is no emission. The yellow zone denotes the standard deviation of the frequency. (f) Schematic of a magnetic tunnel junction as a synapse, with an SEM image of the device. The yellow strip is a field line that allows generating a local DC magnetic field. (g) Schematic of two magnetic tunnel junctions in series forming a synaptic chain. (h) Schematic of a magnetic tunnel junction as a neuron.

Synapses are realized by junctions used as resonators that transform RF to DC[14,31]. A rectified DC voltage is indeed generated when an RF signal is injected into a magnetic tunnel junction within a few percent of its resonance frequency, through the mixing of the induced magneto-resistance oscillations with the input signal[32]. This DC voltage is proportional to the input RF power, as experimentally shown in Fig. 1b. Under these conditions, the magnetic tunnel junction multiplies the RF input by a constant that can be equated to a synaptic weight[31]. The latter depends on the difference between the frequency of the input signal and the resonant frequency of the junction. Fig. 1c shows that the synaptic weight can be tuned by changing the resonance frequency of the junction, which we do here by modifying the local magnetic field via current injection into a metal strip above the device, that we call "field line" (see sketch in Fig. 1f and Methods). By connecting the junctions in series, the rectified voltages they produce add up (Fig. 1g). Fig. 1d shows the total rectified voltage produced by two junctions connected in series when the frequency of the injected RF signal is swept. Since the junctions have different resonant frequencies, they rectify the input signal one after the other, as indicated by the two successive undulations of the measured curve. Such a system can thus perform a weighted sum on simultaneously injected RF signals, where each signal has a frequency close to that of one of the junctions[14,31].

In a neural network, a neuron should perform such a weighted sum on its inputs and apply a nonlinear activation function to the result. As the weighted sum occurs naturally in our circuit by connecting the synaptic devices in series, the only features that are needed for the neurons are a nonlinearity and a conversion of the result back to RF, so that the output of the neuron can be used as an input to other synapses. For this purpose, following Refs. [16,26,27], we emulate neurons with magnetic tunnel junctions used as oscillators. When a DC current is injected in a magnetic tunnel junction, it induces, through spin-torque, magneto-resistance oscillations above a current threshold[1]. The top panel of Fig. 1e shows that the resulting experimental RF power versus DC input curve is non-linear, and strongly resembles the rectified-linear function frequently used as the neuron activation function in deep learning: zero below a threshold, then proportional to the input above the threshold. The magnetic tunnel junction can thus be seen as a neuron that takes a DC input, applies a non-linear activation on it, and then broadcasts the result of the computation as an RF signal. The frequency of RF emissions by magnetic tunnel junctions usually depends on the value of the input current due to the anisotropic landscape of magnetization orbits[33]. Here we clamp the frequency of RF emission by applying a local magnetic field through a field line (Fig. 1h), as shown experimentally in the bottom panel of Fig. 1e (see Methods). This clamping of frequency ensures that the RF oscillations of neurons are always tuned to the resonant frequency range of their corresponding synapses.

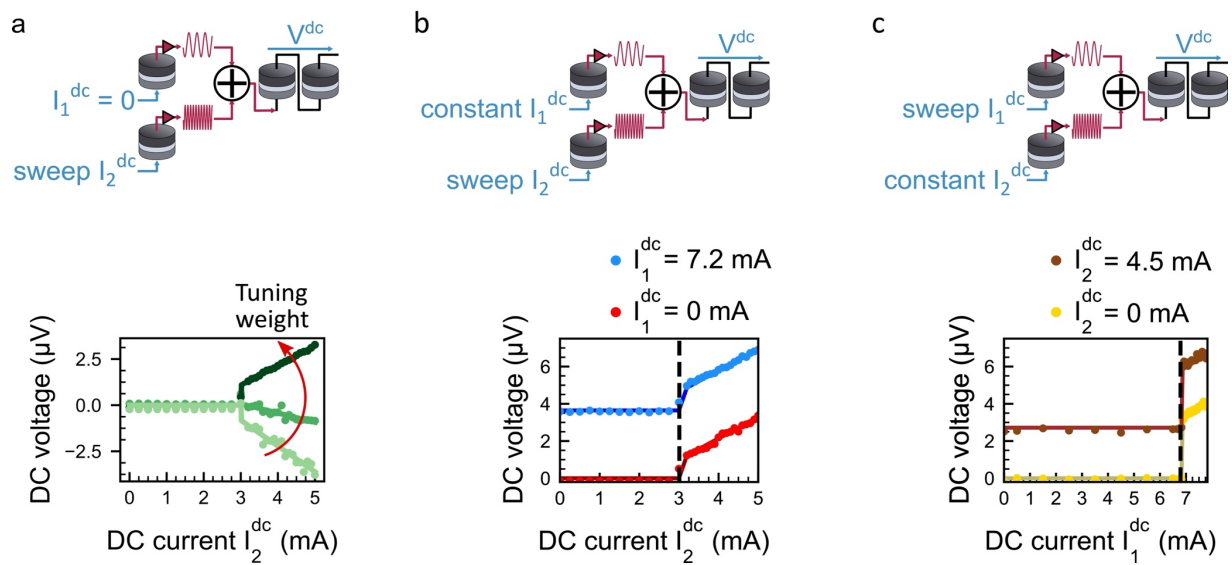

**Figure 2: Experimental demonstration of neurons to synapses interconnection** (a) Connection of neuron 2 to the synaptic chain of two junctions. The dc voltage across the chain of synaptic junctions is plotted versus the dc current in neuron 2, for different currents in the field lines controlling the weights of the synaptic junctions ($I_{FL1}$ is set to 16 mA, and $I_{FL2}$ varies from dark to light green: $I_{FL2}$ = -15, -17 and -18 mA). (b) The input of neuron 2 is swept while the input of neuron 1 is kept constant at 0 (red) and 7.2 mA (blue). (c) The input of neuron 1 is swept while the input of neuron 2 is kept constant at 0 (yellow) and 4.5 mA (brown). For (b) and (c), the dc currents in the field lines are $I_{FL1}$ = 19 mA and $I_{FL2}$ = -18 mA. The two neurons and the chain are measured at perpendicular magnetic fields of 270 mT, 690 mT and 193 mT respectively.

We now employ these building blocks to perform an experimental demonstration of the basic RF/DC interlayer interconnectivity scheme highlighted by yellow dashed lines in Fig. 1a. We use the two junctions of Fig. 1d as synapses, and two other junctions as RF emitting neurons with fixed matching frequencies $F_1$ = 268 MHz and $F_2$ = 383.4 MHz. The synapse junctions are connected in series and receive the RF outputs of both neurons, amplified then summed with a power combiner (see Methods and sketches in Fig 2a-c). We first study the connection between an input neuron and its matching synapse in the chain as illustrated in Fig 2a. The bottom panel displays the evolution of the total voltage $V_{dc}$ across the synaptic junctions as a function of the DC current $I_2^{dc}$ at the input of neuron 2, while the input $I_1^{dc}$ of neuron 1 is set to zero. The measured voltage becomes non-zero above a certain DC current corresponding to the RF emission threshold of neuron 2, here 3 mA. The different curves are obtained by setting different local magnetic fields through the field line above synapse junction 2. They show that the synaptic junction 2 effectively multiplies the activation function of neuron 2 by a weight that can be tuned from positive to negative values by changing the resonant frequency of the synapse.

We then show that the two input neurons actively contribute to the output of the synaptic chain. Fig. 2b displays similar measurements to Fig. 2a, but this time the synaptic weights are kept fixed, and two cases are compared: $I_1^{dc}$ = 0 and $I_1^{dc}$ = 7.2 mA. For $I_1^{dc}$ = 0, neuron 1 does not emit any RF signal and does not contribute to the measured rectified voltage. In contrast, for $I_1^{dc}$ = 7.2 mA, neuron 1 emits a constant RF signal, which induces a constant shift in the total rectified voltage measured. Fig. 2c shows the same results when the roles of the two neurons are reversed. The rectified voltage increases above ~7 mA which corresponds to the RF emission threshold for neuron 1.

In the plots of Fig. 2a-c, the dots correspond to the experimental measurements and the lines to the expected neural network behavior modelled through:

$$V_{dc} = a_1\left(I_1^{dc}\right) \times W_1 + a_2\left(I_2^{dc}\right) \times W_2 \quad (1),$$

where $a_1$ and $a_2$ are the activation functions of neurons 1 and 2, and $W_1$ and $W_2$ are the weights of synapses 1 and 2 (see model details in Methods). The excellent agreement between the experimental data and the neural network model in Eq. 1 demonstrates that the magnetic tunnel junctions interconnected through successive conversions between DC and RF implement the neuron and synaptic operations of a neural network in a very clean manner. This result also constitutes the first demonstration of a functional connection of nanoscale neurons to nanoscale synapses. Indeed, although the connection of memristive nano-synapses to nano-neurons has been achieved in several works[34–36], the signals provided by nanoscale memristive neurons tend to exhibit large fluctuations and require reshaping before they can be injected into the memristive synapses.

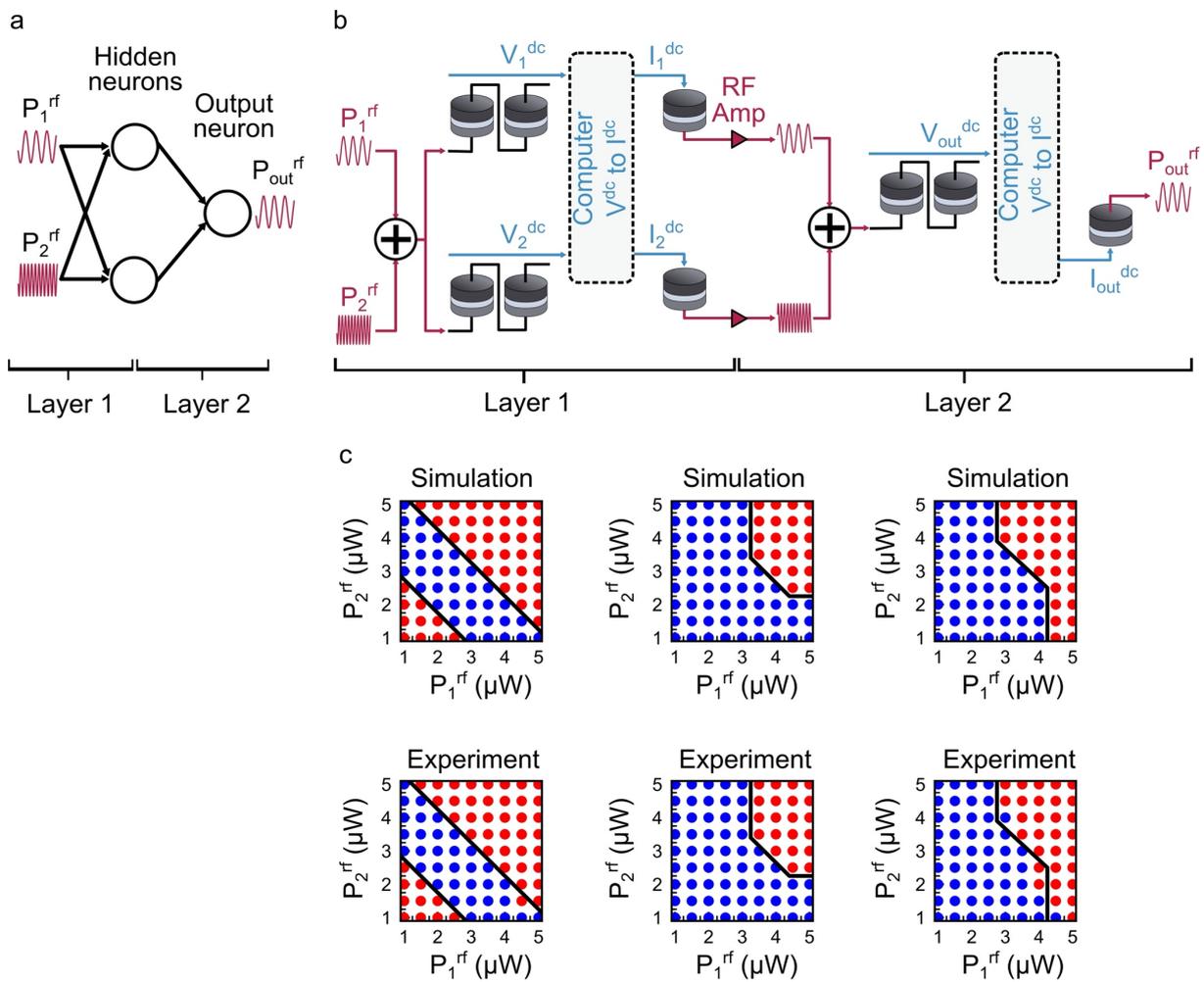

**Figure 3: Experimental demonstration of the spintronic neural network.** (a) Equivalent software network, composed of two RF inputs, two hidden neurons and one output neuron. (b) Experimental implementation composed of nine magnetic tunnel junctions: four of them mimicking synapses, and three neurons. (c) Non-linear RF classification task. Each panel shows a 2D plane where each data point has the coordinates ($P_1^{rf}$, $P_2^{rf}$) and a color corresponding to its output class (red or blue). The black lines are the target boundaries of the classes. The top row shows simulations with the equivalent software network while the bottom row shows experimental results. Each column is a different task. The three chains and three oscillators are measured at perpendicular fields of 230 mT, 600 mT, 193 mT for the chains, 270 mT, 690 mT and 415 mT for the oscillators.

We use this low noise and high-quality connection scheme to build the two-layer neural network illustrated in Fig. 3a, composed of three neurons (two hidden and one output) connected by six synapses. The inputs are composed of two RF signals with fixed frequencies $F_1$ and $F_2$ and varying powers $P_1^{rf}$ and $P_2^{rf}$. The corresponding hardware implementation is shown in Fig. 3b. We measure the DC voltages across the serially connected synapse junctions and feed the neurons with proportional DC currents, using external instruments and a computer interface (see Methods). To probe the performance of the spintronic network, we perform three different non-linear classification tasks illustrated in the top panels of Fig. 3c. The goal is to separate the dots in two classes in the 2D plane ($P_1^{rf}$, $P_2^{rf}$): "red" if the output neuron emits an RF signal ($P_{out} > 0$) and "blue" if it does not ($P_{out} = 0$). We train the network offline using the equivalent software network of Equation 1 with parameters extracted from the experiment for neurons and synapses. We then set the experimental weights as close as possible to the trained software network (see Methods). The bottom panels in Fig.3c show the corresponding experimental results using the offline trained weights and biases (see Methods). The black solid lines delimit the target classification. We observe high accuracies of 98%, 100% and 95% – 97.7% on average – that demonstrate the capacity of the network to perform clean, non-linear classification of RF signals as shown in Fig. 3c. The small drop in accuracy compared to the purely software neural network – that achieves 100% on the three tasks – is due to measurement noise causing some data points close to the decision boundary to be misclassified, which could be optimized for future implementations. The training procedure requires to access a wide range of synaptic weights in hardware, as well as accurate models of the device properties. Here this is possible because of the tunability of spintronic devices, and the excellent fit between model and experiments. The fact that the solved tasks are non-linear (i.e. here the red and blue regions cannot be delimited by a single line) demonstrates the critical role of the hidden neurons and of the inter-layer connection.

A significant advantage of the proposed network architecture is the possibility to process different types of inputs. If the first layer is a neuron layer, it takes DC inputs, which can connect it to standard electronic circuits. If the first layer is a synaptic layer, it can directly process RF inputs as demonstrated with the experimental network. In the latter case, the synaptic layer plays the role of a spectrum analyzer, and the extracted results can then be refined and combined in the next neuron layers to automatically classify the input RF signals. Such native RF classification has a wide range of applications including medicine[37–39], rf fingerprinting[40] and gesture sensing[41].

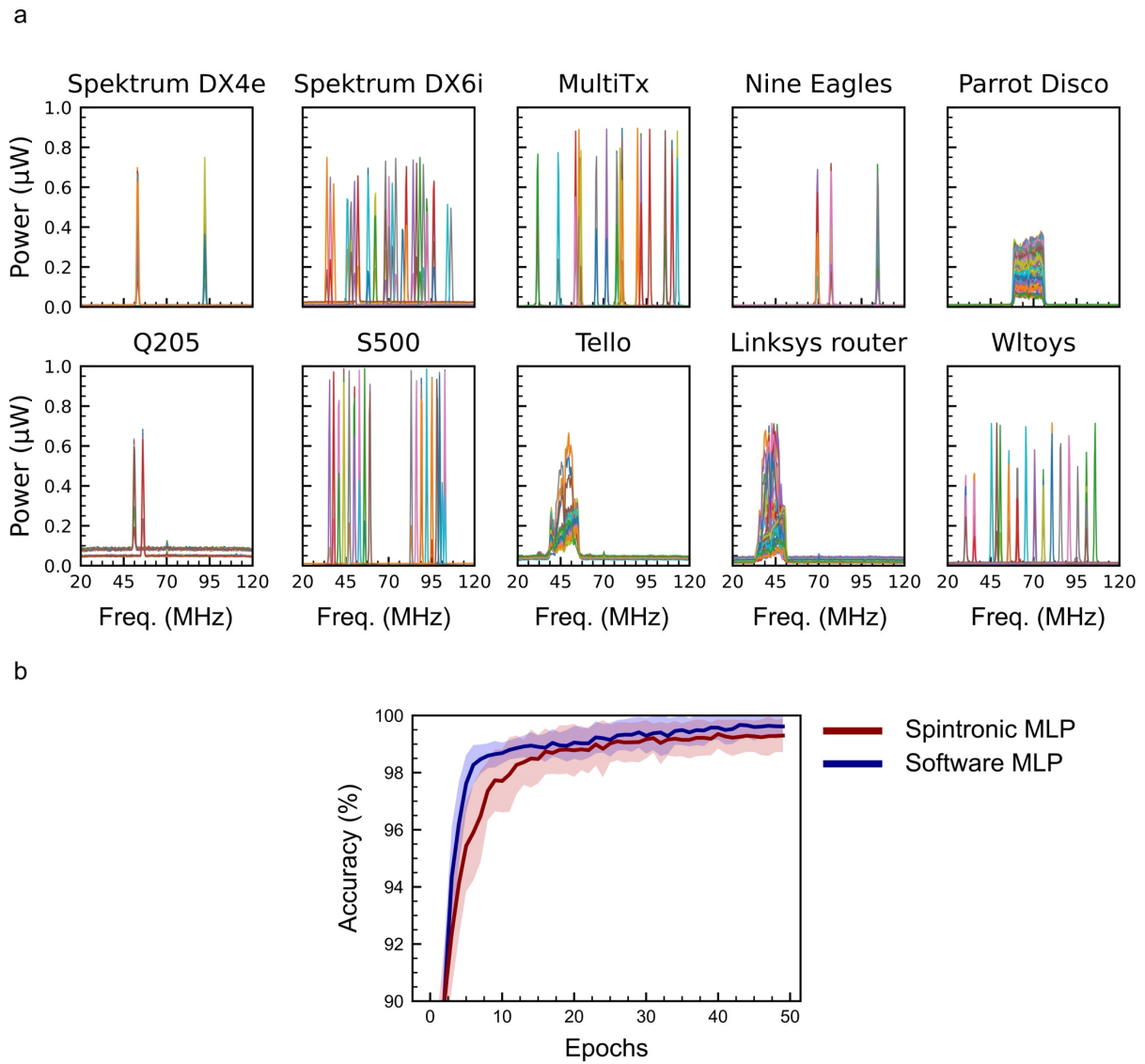

**Figure 4: Drone classification by direct processing of their RF emissions** (a) Spectral representation of the dataset. Each panel represents the power versus frequency for one class of emitter. Each color is one example. (b) Accuracy versus number of training epochs for the simulated spintronic (red) and software (blue) networks. The network architecture is a multilayer perceptron (MLP) with 256 inputs, 128 hidden neurons and 10 outputs.

We show here through simulations that a network of 256 inputs, 128 hidden neurons and 10 outputs can identify different drone types directly through their Wi-Fi emissions. We use the drone dataset of Ref.[42] that considers nine different types of drones, drone controllers and a Wi-Fi router (see Methods). Fig. 4a presents examples of RF signals for the different drone types, that serve as inputs to the network. The synapses in the first layer directly processes the input RF signals and thus are designed to match the input frequencies. In contrast, the frequencies of the synapses in the output layer are chosen to match the frequencies of the neurons in the hidden layer, which can be set independently of the input frequency range. Each chain of synapses receives and processes all output signals from the previous layer. Fig. 4b shows the evolution of the classification accuracy versus the number of training epochs, for the simulated spintronic neural network as well as a pure software network, using backpropagation (see Methods). We reach a maximum accuracy of $99.3 \pm 0.58$ % with the spintronic network and $99.62 \pm 0.39$ % with the software network, where the uncertainties correspond to the standard deviation over a hundred trials. This demonstrates that the spintronic neural network can perform classification tasks on real-world applications with an accuracy equivalent to that of a software neural network. Furthermore, the spintronic network has the considerable benefit to natively process RF inputs, without requiring any digitalization contrarily to software, or other electronic implementations.

The RF spintronic neural network has the significant advantage for applications to process RF inputs natively, without requiring any digitization. Currently, performing the drone classification task in this work would typically require digitizing the RF inputs with an Universal Software Radio Peripheral, a bulky device consuming about 45 W[43], as Basak et al. did[42], and then performing the neural operation on a Graphical Processing Unit that consumes more than a hundred Watts[44]. In contrast, the spintronic neural network simulated in this work, that natively processes RF inputs, would consume only 3.4 mW (see Methods) for full processing, which is a gain of four orders of magnitude over current methods in terms of power consumption.

Indeed, with a diameter of 20 nm for each junction, such a large scale network would consume approximately ten femtojoules per synaptic operation and one hundred femtojoules per neural operation, amplifiers included[45]. This energy consumption takes into account the RF and DC amplifiers needed between each layer of neurons and synapses to maintain signal levels throughout the network. It is comparable or lower to the estimated energy consumption of alternative neuromorphic implementations with memristive or photonic components, and several orders of magnitude less than current implementations of software neural networks (see Methods)[46].

In these future networks, weights can be tuned in a non-volatile and low-power manner by tuning the resonant frequency of synaptic junctions through voltage-controlled magnetic anisotropy effects instead of the DC injection that we use here to generate a magnetic field through the field line[12,25,47]. Similarly, the emission frequency of the neuronal junctions can be fixed by using junctions with a weak magnetic anisotropy, to avoid the DC-current-generated magnetic field employed in our proof of concept[48,49]. Finally, the use of different fields for each device can be avoided in the future by material and nanofabrication optimization to reduce the variability and set the range of operations of the oscillators and resonators at zero field[50,51].

The new RF connection scheme based on frequency multiplexing proposed here also constitutes a promising alternative to the crossbar array geometry for densely connecting neurons through synaptic devices with limited resistance variations. Passive crossbar arrays indeed typically require devices with an OFF/ON ratio above 100 to avoid excessive parasitic currents between their columns and rows[3], and they are not adapted to connect low OFF/ON memristors or magnetic tunnel junctions[52]. In contrast, the presented RF connection scheme does not suffer from sneak paths: by construction, the synaptic chains feeding each output neuron are not interconnected as the communication between neuron and synapses relies on frequency multiplexing instead of pure wiring, as can be seen in in Fig. 1a.

The frequency range accessible to magnetic tunnel junctions and their resonance width sets the maximum number of synaptic junctions in each chain. With a range of 0.05 to 20 GHz[28] and a resonance width of 1% of the resonant frequency, we estimate through simulations that each chain of serially connected junctions can host 784 synapses while maintaining a high recognition accuracy, taking into account the overlap between the resonance bands of different synapses (see Methods). 784 neurons can therefore be connected to 784 other neurons through an all-to-all connection, which is larger than what can be achieved with passive memristive crossbars due to sneak current paths[53]. In addition, the type of connectivity that can be achieved is highly flexible, from fully connected to convolutional neural networks, as presented theoretically in[31,54].

An important step will be to demonstrate that RF signals can be propagated without excessive losses through chains composed of a large number of junctions. Here, the multifunctionality of spintronics will be key to success. The RF signals can indeed be sent directly in the synaptic chains, or in an adjacent stripline electromagnetically coupled to the chain. The junctions can be insulating, with large magneto-resistance and high resistance, or purely metallic, with high conductivity and lower magneto-resistance. Recent results show that up to 64 metallic spintronic junctions can synchronize their RF emissions, which indicates that RF signals can propagate in systems composed of a large number of junctions[21]. Furthermore, current progress in spin-charge conversion and THz spintronics can be exploited in the future to improve the overall performance of the networks in terms of speed and energy consumption[55,56].

In conclusion, we have demonstrated a multilayer spintronic neural network made of magnetic tunnel junctions interconnected through successive conversions between RF and DC. This scheme constitutes a promising alternative to the crossbar geometry for spintronic devices for building low-power, densely interconnected large-scale neural networks. We have shown through simulations that a scaled-up network can natively classify RF signals with orders of magnitude less energy than standard approaches, with important applications at stake in telecommunications, medicine, autonomous vehicles and traffic regulation. The intrinsic dynamics of these networks is moreover an exciting opportunity to train them on chip by algorithms that exploit physical effects for learning[57].

**Methods**

**Magnetic tunnel junction samples preparation**

Magnetic tunnel junctions (MTJs) films have a stacking structure of: Buffer [5 Ta / 50 CuN / 5 Ta / 50 CuN / 5 Ta / 5 Ru] / Antiferromagnet [6 IrMn] / Synthetic Antiferromagnet [2.0 $CoFe_{30}$ / 0.7 Ru / 2.6 $CoFe_{40}B_{20}$] / Tunnel barrier [~1 MgO (RxA product in the 3-12 Ω µm$^2$ range)] / Free layer [2.0 $CoFe_{40}B_{20}$ / 0.2 Ta / 7 NiFe] / 10 Ta / 7 Ru (thicknesses in nm).

The MTJ stack is deposited by magnetron sputtering by the Timaris MTM tool. After an e-beam lithography with a negative ARN7520.18 resist, the pillars are defined by ion beam milling. The free layer is a composite of CoFeB and NiFe. The CoFeB is amorphous as deposited and crystalises into bcc 100 during annealing (330°C, 1 Telsa, 2 hours), which is necessary to ensure high quality MgO growth and the NiFe is a soft magnetic material which dominates the dynamic behaviour. The 0.2 Ta is deposited to stop the NiFe fcc 111 influencing the crystallisation at the CoFeB. The pillars have a nominal diameter of 350 nm.

The free layer has a magnetic vortex as the ground state. In a small region called the vortex core, the magnetization spirals out of plane. Under DC or RF current injection and the action of the Oersted field and spin transfer torques, the core of the vortex gyrates with a frequency in the range of 150 MHz to 450 MHz for the oscillators and resonators we used here.

The typical resistance versus in-plane field profile is presented in Fig. S1. The tunnel magneto-resistance is 72 %.

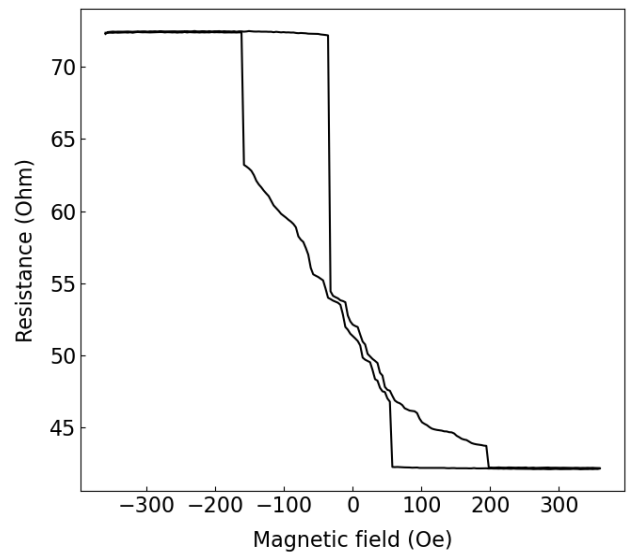

*Figure S1: Resistance versus magnetic field of a magnetic tunnel junction, for a DC current of 0.1 mA.*

In the case of chains of MTJs connected in series, the connection is done by lithography through the top and bottom electrodes of the devices.

**Field line for local magnetic field application**

The field line is a 300 nm thick and 3 $\mu$m wide AlSiCu stripline, deposited 600 nm above the free layer of the MTJ. The local magnetic field generated by DC current injection in this stripline is collinear with the free layer and about 0.05 – 0.1 mT / mA.

**Characterization of the synaptic and neuron devices of the fully spintronic neural network**

**Synaptic chains.** In the network, we have used three chains of two serially-connected synaptic junctions each. Here we show their response to a frequency-swept RF signal. The chains presented Fig. S2 a-c are measured under out-of-plane magnetic fields of 193 mT, 230 mT and 600 mT respectively.

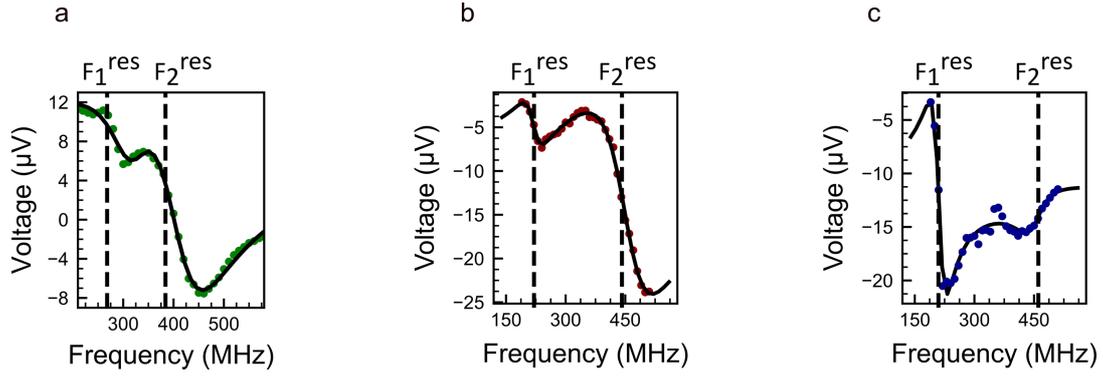

*Figure S2: Spin diode response of the three synaptic chains in the network.* (a) Chain for the second layer of synapses, for an RF input of 7 µW. (b-c) Chains for the first layer of synapses, both for an RF input of 11 µW.

**Neurons**. In the network, we have used three neurons: neurons 1, 2 (hidden) and 3 (output), represented on the three columns of Fig. S3 respectively. The measurements are performed under out-of-plane magnetic fields of 270 mT, 690 mT and 415 mT respectively. Fig. S3 a-c show typical emission spectra of the neurons. The DC currents in the devices are 8 mA, 5 mA and 6 mA respectively. The DC currents in the field lines are 8 mA, 2 mA and 18.5 mA respectively. Fig. S3 d-f show the emitted RF power and frequency versus DC current before clamping the frequency (i.e. constant current in the field line). Fig. S3 g-i show the emission RF power and frequency versus DC current after controlling the DC current in the field line ($I_{FL}$) in order to maintain a constant emission frequency. The frequencies of the neurons 1, 2 and 3 are fixed at 266.46 ± 1.36 MHz, 384.09 ± 0.90 MHz and 254.68 ± 0.19 MHz respectively. The value of the DC current to inject into the field line ($I_{FL}$) at different values of dc current into the device ($I^{dc}$) is found through prior characterization and interpolation. These measurements are done with no amplification for neurons 1 and 2 and with 11 dB amplification for neuron 3.

**Interconnection between two neurons and two synapses.** Neuron 1, neuron 2 and the chain of synapses are measured in the same conditions as their characterization described above. We use a dedicated experimental setup that enables us to apply different magnetic fields to different devices simultaneously, using pairs of permanent magnets. Each of the parts (the two neurons and the chain) is connected to SMA connectors through wire-bonding. Then each neuron is connected to a circulator (to protect the device from reflections), then to an RF amplifier. The amplifiers are of 6 dB and 11 dB for neurons 1 and 2 respectively. The amplifications were chosen such that the output powers of the two hidden layer neurons were comparable in the final network (maximum powers of 3.4 µW and 3.9 µW respectively). The two resulting signals are combined by a power combiner (6 dB loss) and sent to the chain. DC inputs are provided via a multi-channel SMU, to the neurons and their field lines, as well as to field lines of the synapses in order to vary their weights.

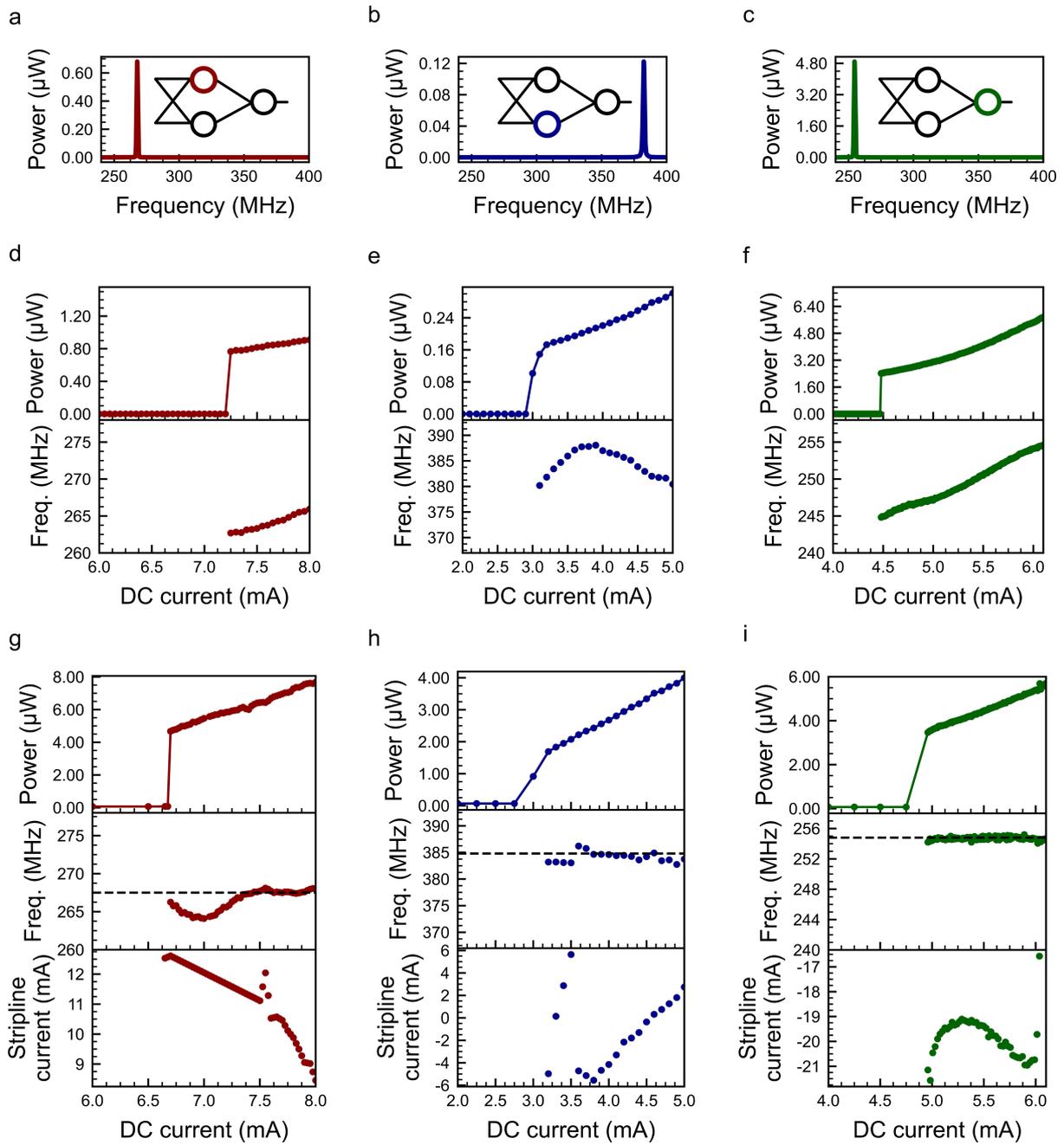

*Figure S3*: **Characterization of the neurons.** *Each column is one neuron. (a-b-c) Power versus frequency spectra. (d-e-f) Power and emission frequency versus DC current in the sample, before fixing the frequency. (g-h-i) Power and emission frequency versus DC current in the sample, after fixing the frequency. DC current in the field line versus dc current in the sample.*

Fig. S4 shows the same measurement as Figure 2 a, but for neuron 1, i.e. the connection of neuron 1 to the synaptic chain, for different synaptic weights (colors). Here the dc current in neuron 2 is null, so that there is no RF output from neuron 2.

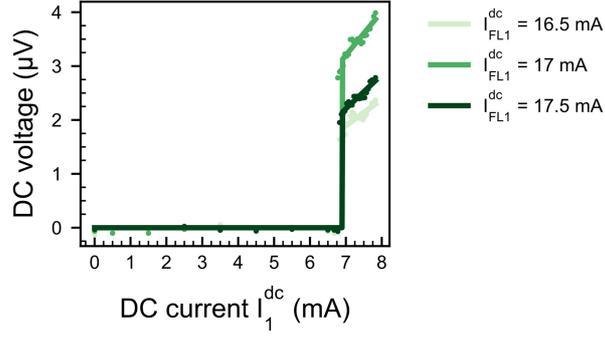

*Figure S4: Connection of neuron 1 to the synaptic chain. DC voltage across the chain versus dc current in the neuron, for different DC currents in the field lines of the synapses (colors). The currents in the field lines to control the weight are, from light to dark green: $I_{FL1}$ = 16.5 mA, 17 mA and 17.5 mA, $I_{FL2}$ = -15.5 mA.*

In total, 49 sets of weights were obtained ($I_{FL1}$ from 16 to 19 mA by 0.5 mA steps, $I_{FL2}$ from 15 to 18 mA by 0.5 mA steps). For each set of weights, the experimental data is compared with the following model:

For each set of weights, the experimental data is compared with the following model:

$$V_{dc} = V_1(I_1^{dc}) + V_2(I_2^{dc})$$

$$\text{With } V_i(I_i^{dc}) = \begin{cases} 0 & if \ I_i^{dc} < I_{th,i} \\ W_i \times I_i^{dc} + c_i & if \ I_i^{dc} \geq I_{th,i} \end{cases} \quad (S1)$$

Where the weights $W_i$, the constants $c_i$ and the threshold current $I_{th,i}$ are constants extracted from the experimental data. $I_i^{dc}$ are the input DC currents to each neuron.

For the combination of weights shown in Figure 2 b and c ($I_{FL1}$ = 19 mA and $I_{FL2}$ = -18 mA), the root-mean-square error over all oscillator currents is 0.17 µV, i.e. 2.3 % of the range. For all 49 measured combinations of weights, the root-mean-square-error is 0.30 µV, i.e. 3.6 % of the range

**Two-layer fully spintronic hardware neural network**

The three synaptic chains and three neurons are measured in the conditions of their characterization described above. The two input RF signals to the network have fixed frequencies of $f_1$ = 220 MHz and $f_2$ = 400 MHz.

The two inputs are combined through a power combiner and the resulting signal is injected in the two synaptic chains of the first layer through a power splitter. The input powers each range from 0.5 to 5 µW. For each input powers pair ($P_1^{rf}$, $P_2^{rf}$), the output DC voltage of each chain is recorded and converted into a DC current by multiplication on the computer. A bias is added to each DC current to mimic the biases of neurons. The resulting DC currents are then injected into neurons 1 and 2, which RF outputs are amplified, combined and injected into the synaptic chain of the second layer (as described in the interconnection Method). The output voltage of the synaptic chain of the second layer is recorded and converted into a DC current by multiplication on the computer. A bias is added. The resulting DC current is then injected into neuron 3, which output RF power is amplified by 11 dB and read as the output of the network.

**Experimental RF signal classification**

The equivalent software neural network is composed of the analytical models following Equation (1) of the synaptic chains (weighted sum), the neuron activation functions, as well as one bias for each neuron.

We define three different non-linear tasks. For each task the procedure is as follows. We define boundaries in the (Power 1, Power 2) 2D-plane of input power values, marked by black lines, that delimit two regions (class 0 and class 1). We train the equivalent software neural network using the neural network library PyTorch through backpropagation (Adam as optimizer and mean-square-error as loss), so that the output of the network is 1 in one region (red) and 0 in the other (blue). Thus, the inputs are (Power 1, Power 2) couples and the targets are 0 or 1. The training provides a set of ideal weights for the synapses. We set the DC currents in the field lines of the synaptic chains so that the synaptic weights match the ideal weights as close as possible. We then perform the experimental classification tasks by injecting (Power 1, Power 2) input pairs into the hardware network.

**Drone classification dataset**

The dataset is composed of several spectrograms of signals of 9 drone radio controllers and additional Wi-Fi signals forming a total of 10 classes of signals. The signals, originally in the WiFi band, were recorded using an USRP with down-conversion to the 0-100 MHz range, by Basak et al.[42]. The spectrograms have a size of 256 by 256 and were averaged over the FFT frame numbers, representing the time, which gives discrete Fourier transforms with an averaged amplitude. This amplitude has been scaled over the whole dataset from 0 to 1 µW. For better matching with the physical network, we translate the signals to the 20-120 MHz range.

**Physical models used for the simulated neural network**

**Weighted sum.** In our case, the input spectrum is composed of a finite number $N_{inputs}$ of equidistant input frequency bins so that the total DC voltage rectified by one resonator of resonance frequency $f^{res}$ is the sum of the individual effects of each frequency:

$$V(f^{res}) = \sum_{i}^{N_{inputs}} P(f_i) G(f_i, f^{res})$$

where $P(f_i)$ is the input power and $G(f_i, f^{res})$ is the rectification coefficient:

$$G(f_i, f^{res}) = \frac{2\alpha f^{res}(f_i - f^{res}) K_{SD}}{(\alpha f^{res})^2 + (f_i - f^{res})^2}$$

where $\alpha = 0.01$ is the Gilbert damping and $K_{SD}$ is the spin-diode sensitivity.

We consider $N_{res}$ resonators per synaptic chain, where each resonator k of a chain j will rectify the input signal. As a result, the rectification function can be written conveniently using a 3-tensor:

$$G_{jik} = \frac{2\alpha f_{jk}^{res}(f_i - f_{jk}^{res}) K_{SD}}{\left(\alpha f_{jk}^{res}\right)^2 + \left(f_i - f_{jk}^{res}\right)^2} \ .$$

The weight applied by the synaptic chain j to the input $P(f_i)$ corresponds to the total rectification done by all the resonators of the chain $j$ for the frequency component $f_i$:

$$W_{ji} = \sum_k (-1)^k G_{jik}$$

where $(-1)^k$ comes from the fact that the resonators are electrically connected in a "head-to-head" manner. Indeed, when resonators are connected "head-to-tail" (as shown in Fig. 1g), their low frequency finite component add up and create a voltage drift (as observed in Fig. 1d) that increases with the number of resonators. For chains with a large number of resonators, it is therefore preferable to connect the resonators "head-to-head", as depicted in Fig. 1a.

The output voltages are expressed as a classical neural network weighted sum:

$$V_j = \sum_i P_i W_{ij} + b_j\,.$$

We conveniently write this equation using vectors and matrices

$$\boldsymbol{V} = \boldsymbol{PW} + \boldsymbol{b}\,. \text{ (S2)}$$

The voltage bias $\boldsymbol{b}$ is actually composed of two terms:

$$\boldsymbol{b} = \boldsymbol{V}^{\text{chains}} + \boldsymbol{V}^{\text{layer}}$$

where $\boldsymbol{V}^{\text{chains}}$ is the vector containing the voltage biases of the chains (learning parameters) and $\boldsymbol{V}^{\text{layer}}$ is a vector of the same dimension whose components correspond to an additional constant voltage bias (hyperparameter) used to improve the performance of the network. This constitutes the first operation of the network, i.e. the weighted sum.

For the numerical values, N is fixed by the dataset to 256, $\alpha = 0.01$ and $K_{SD} \approx 8.8 \times 10^3$ µV/µW.

**Activation function.** The activation function is modelled by the response of a spin-torque nano-oscillator (STNO) to a DC current $I_{\text{DC}}$. If this current is superior to a threshold current $I_{\text{th}}$ the oscillator enters into a stationary precession regime and its normalized magnetization oscillation power can be modelled by:

$$p = |c|^2 = \begin{cases} \dfrac{\xi - 1}{\xi + Q} & if\ I_{\text{DC}} > I_{\text{th}} \\ 0 & if\ I_{\text{DC}} \le I_{\text{th}} \end{cases}$$

where $p$ and c are the normalized power and amplitude of the stationary precession, $\xi = I_{DC}/I_{th}$ and $Q = 2$ is the non-linear damping coefficient. The resulting power of the $j^{th}$ oscillator is:

$$P_j = A\, p(I_{\text{DC}}) \left(\frac{\Delta R}{R} \beta_s\right)^2 R\, I_{DC}^2,$$

where A is a scaling parameter, $R$ is the resistance, $\frac{\Delta R}{R}$ is the tunnel magneto-resistance and $\beta_s$ is the shape factor. We simplify this expression by considering $\frac{\Delta R}{R}\beta_s \sim 1$

$$P_j = A\, p(I_{\text{DC}}) R I_{DC}^2.$$

Additionally, the input current of each oscillator of the hidden layer is clamped $I_c = 4I_{th}$ to mimic the security preventing the oscillators to break. The resulting output powers of the hidden layer are therefore

$$P_j^{\text{hidden}} = \begin{cases} A\, p(I_c) R I_C{}^2 & if\ I_{DC} \geq I_c \\ A\, p(I_{DC}) R I_{DC}{}^2 & if\ I_c > I_{DC} > I_{th} \\ 0 & if\ I_{DC} \leq I_{th} \end{cases}$$

We chose $A = 1.25$ to obtain maximum output power $P$ to 1 µW, $I_{th} = 10$ µA and $R = 1$ kΩ. These values correspond to an estimation of what could be achieved with MTJs with a diameter of 35 nm[45].

In order to obtain the vector of currents $\boldsymbol{I}_{DC}$ from the output voltages obtained with Equation (S2), we are using a voltage-to-current ratio $g_m$ mimicking a transconductance amplifier:

$$\boldsymbol{I}_{DC} = \boldsymbol{V} g_m$$

These previous equations form the building blocks of the inference of the physical deep neural network.

**Neural network architecture.** We consider the same architectures for the equivalent software and simulated physical neural networks: a two-layer neural network with 256 inputs (corresponding to the 256 frequency bins from 20 to 120 MHz), 128 hidden neurons and 10 outputs (corresponding to the 10 classes).

While for the equivalent software neural network the trained parameters are the weights and biases, for the simulated physical neural network the trained parameters are the resonance frequencies of all the resonators $f_{jk}^{\text{res}}$ and the output voltage biases $V_j^{\text{chains}}$.

**Model selection.** For the model selection, the dataset was split into the default train/test datasets, containing 702 and 298 signals respectively, following the original publication of the dataset. For training we used the Cross-Entropy loss and the Adam optimizer.

The hyperparameters of each final model were selected with the use of a hyperparameter optimization procedure based on the Optuna framework. For the software model, there is only one hyperparameter which is the learning rate. For the physical model, the hyperparameters are the learning rate as well as the additive bias $\boldsymbol{V}^{\text{layer}}$ on the output voltages of the first synaptic layer and the voltage-to-current ratio $g_m$ converting the output voltages of the first synaptic layer into currents.

The number of trials (i.e. set of hyperparameters tested) was set to 100. The number of epochs was set to 50 and the training/validation procedure was repeated 10 times in order to use the resulting mean accuracy as an objective function for the Optuna optimizer.

For the software model, the resulting learning rate was $\eta \approx 0.06$.

For the physical model, the resulting learning rate was $\eta \approx 1.07 \times 10^{-5}$, the voltage-to-current ratio about $g_m \approx 1.81$ µA/mV and the additive voltage bias $V^{layer} \approx 0.013\ V$.

**Model evaluation.** After the model selection procedure, we use the resulting hyperparameters and use them in new instances of both the software and the physical models. The number of epochs is 50 and the number of evaluations is 100.

**Impact of the frequency range on neural network accuracy**

We demonstrate here that with a resonance frequency range between 50 MHz and 5 GHz and resonance widths of 1% of the resonance frequencies, each chain of synapses can host at least 784 spintronic resonators while performing accurate synaptic operations. For this purpose, we simulate a network made of a single layer of synapses and we use the MNIST (Mixed National Institute of Standards and Technology) dataset made of 784 pixels pictures. Our simulated network is made of ten chains of 784 spintronic resonators in series, which corresponds to a neural network with 784 inputs and 10 outputs. We used 60,000 images to train the network with a learning rate of $\eta = 10^{-6}$ and 10,000 images to test the network. We used the optimizer Adam and the Cross-entropy loss.

We used the expressions described in the "Weighted sum" Method section to simulate the layer of spintronic resonators. To study the impact of the frequency range on the accuracy of the network, we train networks using the same minimum resonance frequency $f_{min}$ = 50 MHz for each chain and different maximum frequencies $f_{max}$: 100 MHz, 500 MHz, 1 GHz, 5 GHz, 10 GHz and 20 GHz. Each experiment is repeated 10 times to gather statistics. In Fig. S5 a we plot the accuracy on the test and train sets after training over 20 epochs. We compare the accuracy of the neural network with spintronic resonators with the accuracy of an equivalent software neural network. The mean accuracy of the network with maximum resonance frequencies equal or above 5 GHz is 91.52%, which is comparable to the mean accuracy of the software network (91.71%), and higher than the mean accuracy of the software network minus its standard deviation (0.26%).

We observe that the accuracy drops for smaller resonance frequency ranges ($f_{max}$ = 100 MHz, 500 MHz and 1 GHz) for both train and test sets. To understand this behavior, in Fig. S5 b-e we plot a subset of synaptic weights corresponding to synaptic connections between four of the ten outputs and 60 of the 784 inputs. We chose synaptic connections from the 400th to the 460th input so that these synaptic connections correspond neither to the beginning nor to the end of each chain of resonators. These synaptic weights are taken from networks trained on one epoch. In Fig. S5 d we see that with a very wide range of frequencies (between 50 MHz and 20 GHz), there is no apparent correlation between neighboring synaptic weights of the same synaptic chain. A spintronic network with such wide range of frequency is ideal in the sense that it is similar to the software network (see Fig. S5 e) and yields high accuracy (see Fig. S5 a). However, with small ranges of frequencies ($f_{max}$= 100 MHz and 500 MHz, see Fig. S5 b and c), we observe that neighboring synaptic weights of the same synaptic chain are very close to each other. This is due to the fact that with these small ranges of frequencies, the separations between the frequencies of the RF input signals are too small compared to the resonance width of the resonators. Hence, different neighboring input RF signals are rectified equally by the chains of resonators. The synaptic weights are then not entirely separable (see Fig. S5 b and c), leading to smaller effective number of synaptic weights, explaining the accuracy drop in Fig. S5 a for small values of the maximum frequency in the chain.

In conclusion, the number of synaptic junctions in a chain is limited by the step between different input RF signal frequencies; if it is too small compared to the resonance linewidths, the network loses its accuracy. We find that with a range of input frequencies between 50 MHz and 5 GHz, each chain can host at least 784 spintronic resonators.

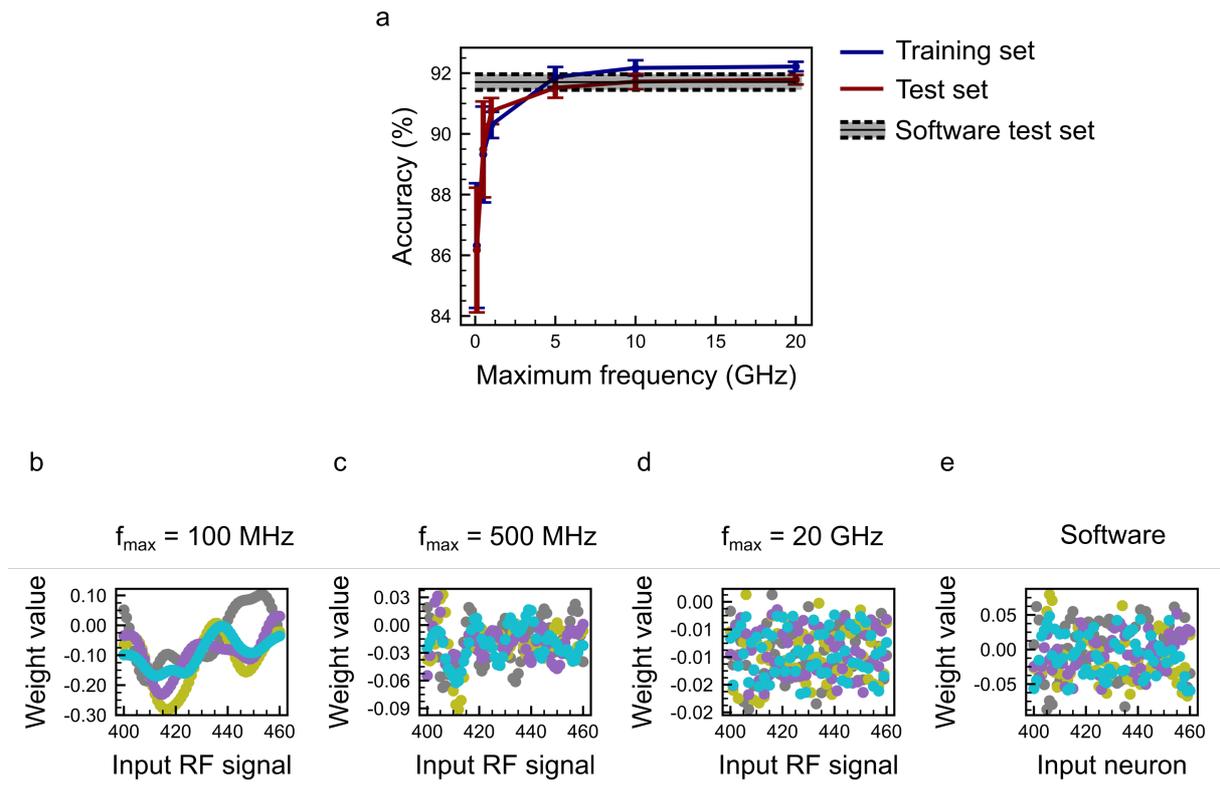

*Figure S5: Demonstration of the accuracy of the spintronic neural network with different range of frequencies. (a) Accuracy of the spintronic network on the test (red) and train (blue) sets of MNIST. Error bars corresponds to the standard deviation. The solid black line represents the mean accuracy of the software neural network on the test set. The upper and lower limit of the grey area respectively represent the mean accuracy of the software network plus and minus its standard deviation. (b-e) Synaptic weights corresponding to synaptic connections between four of the ten outputs and 60 of the 784 inputs. The four different colors (cyan, magenta, grey and olive) correspond to four outputs. (b-d) correspond to spintronic networks with different ranges of resonance frequency, (e) to the software network.*

**Energy Calculation**

A full network can be broken into subsets as shown in Fig. S6, consisting of inputs coming from a layer of N oscillators which are amplified and passed to M chains of N diodes which perform the synaptic operation. Each chain output is then passed to an oscillator in the next layer. We have two layers of amplifiers, the first is an RF amplifier between the N oscillators and M chains and then a second layer of DC amplifiers between the M chains and M oscillators of the next layer.

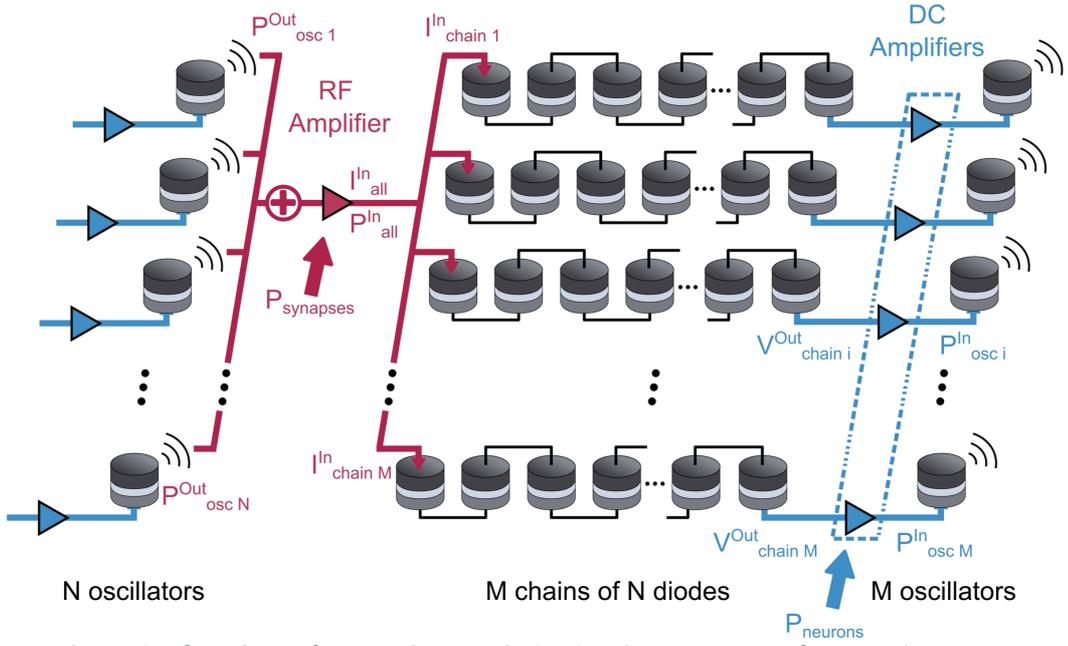

*Figure S6: Schematic of a subset of a neural network circuit*, where N neurons from one layer are connected to M neurons of the next layer.

The peak DC voltage generated by one synapse (i) when receiving input from one neuron (j) is:

$$V_{ij}^{Out} = K_{SD} \times R \times \left(I_{chain\ j}^{IN}\right)^2 \quad (S3)$$

where $K_{SD}$ is the spin diode sensitivity (in µV/µW) and R is the electrical resistance value of a single magnetic tunnel junction. We suppose all devices to have equal resistance for simplicity sake. This then determines $I_{chain j}^{IN}$ to be,

$$I_{chain j}^{In} = \sqrt{\frac{V_{ij}^{Out}}{K_{SD} \times R}} = \sqrt{\frac{V_{min}}{N \times K_{SD} \times R}} = \frac{I_{all}^{In}}{M} \quad (S4)$$

where $I_{all}^{In}$ is the total input current for M chains in parallel and $V_{min}$ is the minimum desired voltage. For N diodes in series and M chains in parallel, the total resistance of the synaptic layer is $R_{all\ chains} = \frac{R \times N}{M}$. The power supplied to the synaptic layer by the RF amplifier, for a single neuron is:

$$P_{all}^{In} = \frac{R \times N}{M} \times \left(I_{all}^{In}\right)^2 = \frac{R \times N}{M} \times M^2 \frac{V_{min}}{N \times K_{SD} \times R} = \frac{M \times V_{min}}{K_{SD}} \quad (S5)$$

The total power that the RF amplifier needs to provide for the total of all neurons is:

$$P_{synapses} = \frac{M \times N \times V_{min}}{K_{SD}} \quad (S6)$$

We can make an estimate for $P_{synapses}$ by noting that the spin diode efficiency can reach $K_{SD} \sim 10^4$ µV/µW and the typical required voltage to feed a DC amplifier and obtain enough precision is $V_{min} = 10^{-3}$ V.

Then, for one synapse:

$P_{synapses} = 10^{-7} W$, i.e. 0.1 µW per synapse.

We can also determine the power supplied to the post-synaptic layer neurons. The power passed to the next layer of oscillators P$_{neurons}$ is given by:

$$P_{neurons} = M \times P_{osc}^{In} = M \times R \times (a \times I_{th})^2 \quad (S7)$$

where $P_{osc}^{IN}$ is the average input power of each neuron, R is the neuron resistance (here we take 2.5 kΩ, which corresponds to a diameter of 20 nm for a resistance-area product RA = 1 Ω. µm²), $I_{th}$ is the threshold current (~$10^{-5}$ A) and a = 2 sets the range of operation of the neuron.

We can then estimate the power consumption of one neuron: $P_{neuron}$ = $10^{-6}$ W, i.e. 1 µW per neuron.

The duration of an operation is linked to the frequency of the devices. For a damping of 0.01, about 100 periods are needed for the steady state to be reached. Assuming devices operating at 1 GHz, the time of operation is then 100 ns.

In consequence, the energy consumption of the operations are:

$$E_{synapse} = 10 \, fJ$$

$$E_{neuron} = 100 \, fJ$$

**Acknowledgements**

This work was supported by the European Union's Horizon 2020 research and innovation program under grant RadioSpin No 101017098.

**Author contributions**

The study was designed and supervised by JG and AM, samples were optimized and fabricated by RF, ASJ, LM, LB, MSC, PA and AJ, experiments and training of the hardware neural network were performed by AM, AR, NL, DM, JT and DS-H, simulations of neural networks and drone classification were performed by AM, AdR, and NL, the estimation of scaled network performances were performed by AM, NL and AFV, all authors contributed to the analysis of results and writing of paper.